\documentstyle[editedvolume]{crckapb} 

\def\gs{\mathrel{\raise1.16pt\hbox{$>$}\kern-7.0pt
\lower3.06pt\hbox{{$\scriptstyle \sim$}}}}
\def\ls{\mathrel{\raise1.16pt\hbox{$<$}\kern-7.0pt
\lower3.06pt\hbox{{$\scriptstyle \sim$}}}}

\def\mpc{\,h^{-1}{\rm Mpc}}

\def\mpch{\,h^{-1}{\rm {Mpc}}}

\begin{opening}
\title{
MEASURING GALAXY BIAS FROM THE HIGH-ORDER\protect\\ CORRELATION FUNCTIONS}

\author{Y.P. JING}
\institute{Research Center for the Early Universe, University of
       Tokyo\\
 Bunkyo-ku, Tokyo 113, Japan}

\end{opening}

\runningtitle{MEASURING GALAXY BIAS}
\begin{document}

\begin{abstract}
  In this talk, I will show how to determine the biasing factor $b$
  from the high-order moments of galaxies. The determination is based
  on the analytical modeling of primordial peaks and virialized halos
  and is independent of the currently unknown density parameter
  $\Omega_0$ and other cosmological parameters. The observed high-oder
  moments of the APM galaxies require that the biasing factor $b$ be
  very close to 1, i.e. the optical galaxies are an unbiased tracer of
  the underlying mass distribution (on quasilinear scale).  The
  theoretical argument can be easily generalized to the three-point
  correlation function and the bispectrum both of which can used as
  further observational tests to the important conclusion of $b\approx
  1$ drawn from the high-order moments. Finally I present our
  preliminary results of the three-point correlation functions for the
  Las Campanas Redshift Survey.

\end{abstract}

\section {Introduction}

A fundamental problem in cosmology is to understand how the spatial
distribution of galaxies (and of galaxy clusters) is related to that
of the underlying mass. The possibility for a galaxy bias
(i.e. galaxies do not faithfully trace the underlying matter in
spatial distribution) is advocated both by observations and by
theories. It is therefore very important to determine the relation
between the distribution of galaxies and that of dark matter (DM). The
distribution of DM cannot be directly measured by telescopes 
because of its very nature. People attempted to understand the DM
distribution by studying the dynamics of galaxies since the peculiar
motions of galaxies are closely related to the distribution of
DM. Unfortunately, these studies can only yield a combination of
$\Omega_0^{0.6} \sigma_8$ where $\Omega_0$ is the density parameter of
the Universe and $\sigma_8$ is the rms tophat density contrast of
radius $8\mpc$. Moreover, there are considerable discrepancies in the
$\Omega_0^{0.6} \sigma_8$ value among different measurements.  In this
talk, I will show how to determine the biasing factor $b$ from the
high-order moments of galaxies, which is based on the analytical
modeling of primordial peaks and virialized halos and is independent of
the currently unknown density parameter $\Omega_0$ and other
cosmological parameters. The method is distinct from the related work
of Gaztanaga \& Frieman (1994) which is based on a simple non-physical
parameterization of the bias relation.

\section{Measuring the biasing factor $b$}

The peak model and the halo model are generally used to prescribe the
formation of galaxies. Recently, we have analytically derived the
high-order moments for the primordial peaks and virialized halos (Mo,
Jing, \& White 1997) based on the method of Mo \& White (1996). The high
order moments of the peaks or the halos, e.g. the skewness $S_3(R)$,
the kurtosis $S_4(R)$ and $S_5(R)$, are functions of the biasing
factor $b$, the shape of the linear power spectrum, and the formation
time $z_f$ only.  Our analytical model has been tested against a set
of high-resolution cosmological N-body simulations, with a conclusion
that the model works very well in the quasilinear regime.

In Figure 1, I present our predictions for the high-order moments
$S_{j,g}$ ($j=3,\,4,\,5$) of peaks and halos as a function of the
linear bias parameter $b$. In this plot, a CDM-like power spectrum
with $\Gamma=0.2$ is assumed.  This spectrum is known to be consistent
with the angular correlation function of the APM galaxies (Efstathiou,
Sutherland \& Maddox 1990).  The currently best estimates of the
high-order moments for optical galaxies are those of Gatzanaga (1994)
based on the APM survey, which are plotted on the figure with
error-bars. Recent observations of galaxies at high redshift indicate
that bulk of galaxies is formed at $z_f\gs1$. Figures 1 then implies
that the optical galaxies are little biased relative to the
distribution of underlying mass (i.e. $b\approx 1$). It is interesting
to note that the observations of $S_3$, $S_4$ and $S_5$ for the APM
galaxies are all consistent with the analytical model if $b\approx 1$.

A comparison with the observed moments of galaxy clusters is also
interesting. Clusters of galaxies are generally believed to be halos
formed recently ($z_f=0$) and have a bias factor $b\approx 3\sim
5$. Our model predicts the skewness $S_{3,h}\approx 2$ and the
kurtosis $S_{4,h} \approx 8$ for such halos which are well consistent
with the observations for the Abell clusters (Jing \& Zhang 1989) and
for the APM clusters (Gatzanaga, Croft \& Dalton 1995).

\begin{figure}
\vskip 8cm
\includegraphics{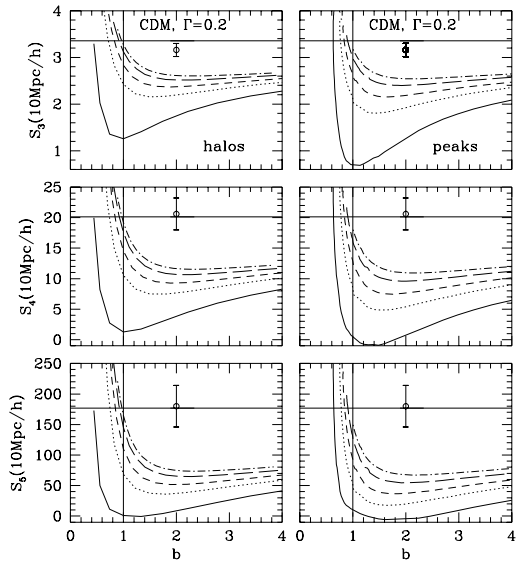}
\caption{
  Model predictions for the high order moments $S_{j,g}$ ($j=3$, 4 and
  5) of halos (left panels) and peaks (right panels) at a radius
  $r=10\mpch$ as a function of the linear bias parameter $b$. Each
  curve shows result for a given formation time $z_f =0$, 0.5, 1, 2
  and 4 (curves from bottom up). The horizontal lines show the values
  of $S_j$ for the mass density field calculated from quasilinear
  theory, whereas the data points (plotted arbitrarily at $b=2$) show
  the observational results for APM galaxies (Gaztanaga 1994)}
\end{figure}

\section{ Further observational tests: the three-point correlation
  function and the bispectrum} 

With a simple extension of the argument of Mo et al. (1997), we can
derive the bispectrum and the three-point correlation function for the
DM halos and the primordial peaks. In this extension we
assume the bias relation by derived Mo et al. (1997) holds at all
scales. We have tested this extension against a set of high-resolution N-body
simulations with $1.7\times 10^7$ particles, and have found that the
extension is very successful (Jing 1997, in preparation). Figure 2
illustrates such a test, where we compare the normalized
bispectrum of N-body dark matter halos (symbols) with the predictions
of the extension. The extension predicts that the optical galaxies
should closely follow the underlying dark matter in the bispectrum and
in the three-point correlation function, if the biasing factor of the
galaxies is about 1 as the high-order moments indicate (\S 2) and the
bulk of the galaxies is formed before redshift 1. Both the amplitude
and the shape dependence of the three-point correlation function and
the bispectrum are potentially useful for constraining the bias factor and the
linear power spectrum of the cosmological models (Fry 1994; Jing \& 
B\"orner 1997; Matsubara
\& Suto 1994; Matarrese et al. 1997).

\begin{figure}
\vskip 9.5cm
\includegraphics{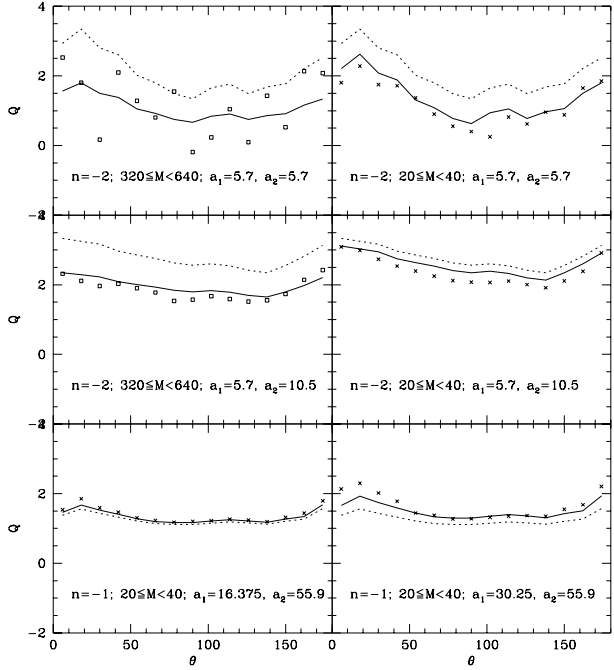}
\caption{The normalized bispectrum $Q(k_1,k_2,\theta)$ of virliazed
  halos estimated from the scale-free simulations with the power
  spectrum index $n$ and $256^3$ particles (symbols), compared with
  the analytical predictions discussed in the text. The halos of mass
  $M$ (in unit of the particle mass) are formed at the scale factor
  $a_1$ and the clustering analysis is made at $a_2$.}
\end{figure}  

\section{Preliminary results from the LCRS survey}

Here I present the preliminary results of our determination (Jing \&
B\"orner 1997) of the three-point correlation function for the Las
Campanas Redshift Survey. With this large sample we are able to
determine the three-point correlation function of galaxies much more
accurately than any previous works (e.g. Efstathiou \& Jedrzejewski
1984). The three-point correlation function is not only a powerful
test to the bias problem as described above but also a generally
important statistic to describe spatial distributions and dynamics of
galaxies (Peebles 1980; Suto 1993).

We analysed two types of three-point correlations: the redshift space
three-point correlation function $\zeta (s,u,v)$ and the projected
three-point correlation function $\Pi (r_p,u,v)$. Our results are
presented in Figure 3. The figure shows that the redshift three-point
correlation function is approximately hierarchical with the amplitude
$Q({\rm red})\approx 0.6$ (see e.g. Jing \& B\"orner 1997 for its
definition).  There is a tendency that $Q$ increases with $v$
especially for large triangles with $s>5\mpc$, qualitatively in
agreement with the model predictions for the mass three-point
correlation function (Jing \& B\"orner 1997). Similar properties are
found for the projected one. However the projected $Q({\rm proj})$ is
about 1.5 for very small scales $r_p\le 0.5\mpc$ and decreases with
$r_p$ for larger $r_p$. If the three-point correlation function in
real space is hierarchical, the projected one should also be
hierarchical with $Q({\rm proj}) =Q$. Our results then support that
the three-point correlation function on small scales is hierarchical
with $Q\approx 1.5$, consistent with the conclusions of Peebles (1980)
based on 2-D angular catalogues. The hierarchical form however seems
not to hold for scales much larger than $0.5\mpc$ because $Q({\rm
  proj})$ decreases with $r_p$ and increase with $v$ for
$r_p>0.5\mpc$.  A quantitative comparison between the observations and
the CDM models is in progress, which will take account of the redshift
distortion effect and the projection effect (Jing \& B\"orner 1997).

\begin{figure}
\vskip9.0cm
\includegraphics{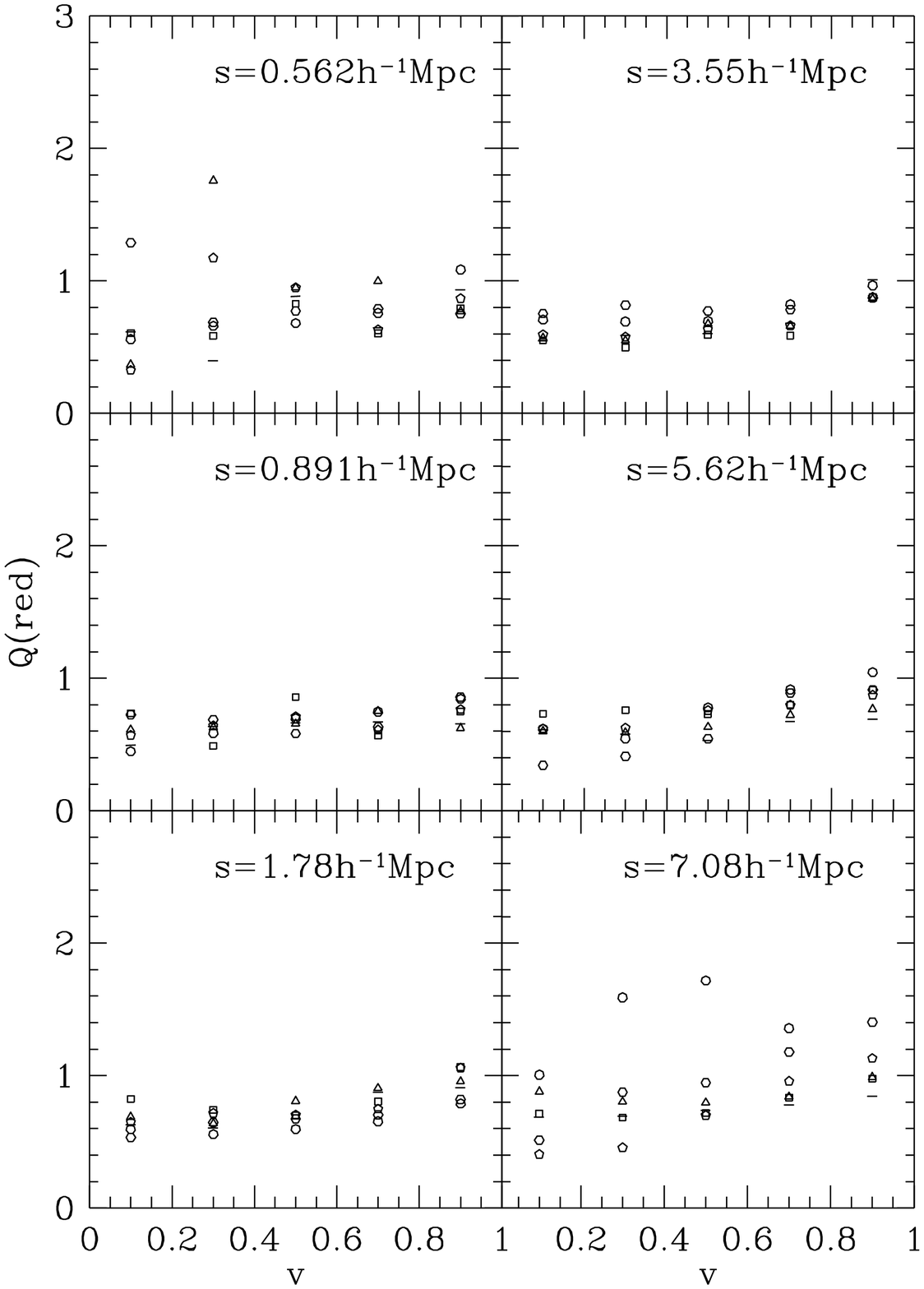}
\includegraphics{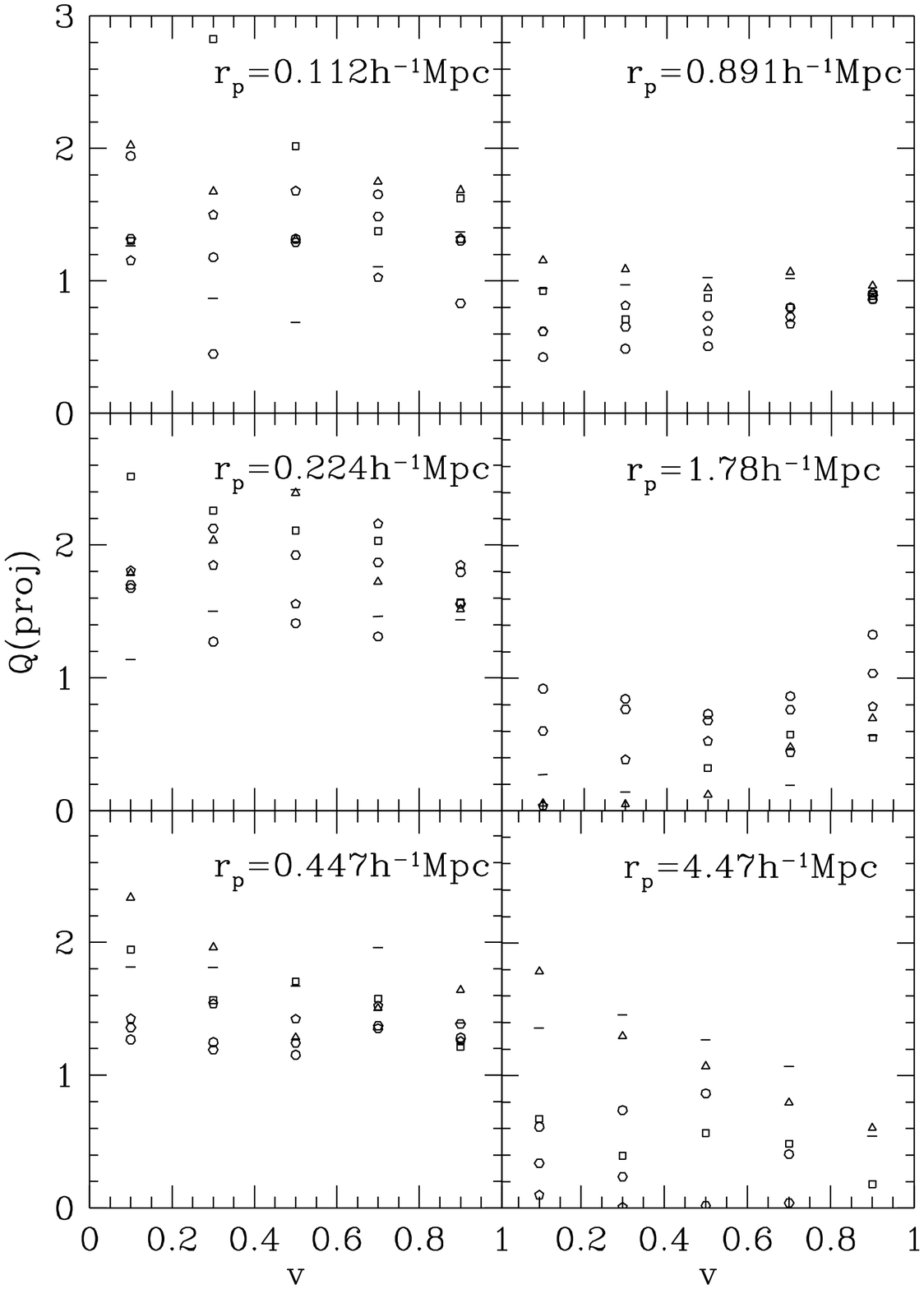}

\caption{The normalized three-point correlation function of
  galaxies in the Las Campanas Redshift Survey. The left panel shows
  the result in the redshift space and the right panel shows the
  projected one. The $r$, $u$, $v$ convention of Peebles (1980) is
  adopted in the plot, and different symbols are for different values
  of $u$. More detailed account of the figure will be given in Jing \&
  B\"orner (1997)}
\end{figure}  

\section{Conclusions}
In this talk I have presented our analytical model for calculating the
high-order moments of peaks and halos in the quasilinear regime.  With
this model, the observed skewness and kurtosis of the Abell and the
APM clusters are interpreted as a consequence of these clusters being
recently formed massive halos. The high-oder moments of the APM
galaxies are consistent with the model predictions if the
galaxies are an unbiased tracer of the underlying mass distribution
(on quasilinear scale).  This model can even be extended to the normal
three-point correlation function and the bispectrum (cf. Catelan et
al. 1997), and thus our important conclusion that $b\approx 1$ can be
further tested with these functions. Finally I presented
the preliminary results of our determination of the three-point
correlation functions for galaxies in the Las Campanas Redshift
Survey.

 \bigskip 
\begin{center}
  {\it Acknowledgements} 
\end{center}
  The author gratefully acknowledges the
  postdoctoral fellowship from Japan Society for the Promotion of
  Science. Part of numerical computations were carried out on
  VPP300/16R and VX/4R at the Astronomical Data Analysis Center of the
  National Astronomical Observatory, Japan,

\end{document}